%
%
%
\def\inc#1{\hbox{\global \advance#1 1}}
\countdef\eqnr=1    
\eqnr=1           
\def\nexteq{\inc{\eqnr}}    
\def\enr{\number\eqnr\nexteq}   
\def\eq{\eqno(\enr)}   

\countdef\HamEq=221
\countdef\EvolEq=222
\countdef\CorrEq=223
\countdef\EnergyEq=224
\countdef\WeakCorrEq=225
\countdef\DefEq=226
\countdef\SuddenUeq=227
\countdef\fmnEq=228
\countdef\fmnEqTwo=229
\countdef\Ueq=230
\countdef\SteadyEq=231
\countdef\SuddenDeq=232
\countdef\OneModeAdiabEq=233
\countdef\AdiabUeq=234
\countdef\AdiabDeq=235
\countdef\Uodeeq=236
\countdef\FKMeq=237
\countdef\ReductionEq=238
\countdef\FourierReductionEq=239
\countdef\FourierEvolEq=240
\countdef\SepEq=241

\countdef\FourierPowerEq=242
\countdef\LimSupEq=243
\countdef\NormBoundEq=244

\countdef\LiapunovEq=250
\countdef\xiEq=251

%
%
\countdef\refnr=102
\refnr=1

\countdef\Huerta=103
\countdef\FKM=104
\countdef\ZurekHabibPaz=105
\countdef\TegmarkYeh=106
\countdef\KimNoz=107
\countdef\HillaryClinton=108
\countdef\Renyi=109
\countdef\FLO=110
\countdef\Zygmund=111
\def\nextref{\inc{\refnr}} 
\def\rnr{$\number\refnr$\nextref} 
\def\ref{[\rnr]}

\def\pp{\parshape 2 0truecm 17truecm 2truecm 15truecm}
\def\rf#1;#2;#3;#4;#5 {\par\pp{\bf \rnr.} #1, {\it #2}, {\bf #3}, #4 (#5).\par} 
\def\rn#1{\par\pp{\bf \rnr.} #1\par}

\def\ie{{\it i.e.}}

\def\llist{\noindent\parshape 2 0.5cm 15.0cm 1.32cm 14.18cm}

\def\slist{\noindent\parshape 2 0.5cm 15.0cm 1.00cm 14.50cm}

\def\om{\omega_0}
\def\vm{{\bf\mu}}
\def\vz{{\bf z}} \def\vx{{\bf x}} \def\vp{{\bf p}} \def\vq{{\bf q}} 
\def\vk{{\bf k}} 
\def\oz{{\bf\hat z}} \def\op{{\bf\hat p}} \def\oq{{\bf\hat q}}
\def\opH{{\hat H}}
 
\def\diag#1{\>{\rm diag}\{#1\}}
\def\expec#1{\langle#1\rangle}
\def\oh{{1\over 2}}
\def\Aoh{A^{1/2}}

\def\ootp{{1\over 2\pi}}
\def\ipp{\int_{-\pi}^{\pi}}

\def\lth{\lambda^2(\theta)}

\def\sumnoi{\sum_{n=-\infty}^{\infty}}

\def\gh{\hat g}

\def\Aoht{A^{1/2} t}

\def\crr{\cr\noalign{\vskip 4pt}}
\def\as{\quad\hbox{as}\quad}
\def\Wh{\widehat{W}}
\def\fh{\hat{f}}
\def\gh{\hat{g}}
\def\R{{\bf R}}

\def\dk{d^dk}
\def\subsection#1{\bigskip\goodbreak{\bf#1}\bigskip}

\baselineskip12pt
\hoffset0.5truecm
\hsize 15.5truecm


\font\titlefont=cmb10 at 15truept
\font\namefont=cmr12 at 14truept
\font\addrfont=cmti12
\font\rmtwelve=cmr12

\newbox\abstr
\def\abstract#1{\setbox\abstr=
    \vbox{\hsize 5.0truein{\par\noindent#1}}
    \centerline{ABSTRACT} \vskip12pt 
    \hbox to \hsize{\hfill\box\abstr\hfill}}

\def\today{\ifcase\month\or
        January\or February\or March\or April\or May\or June\or
        July\or August\or September\or October\or November\or December\fi
        \space\number\day, \number\year}

\def\author#1{{\namefont\centerline{#1}}}
\def\addr#1{{\addrfont\centerline{#1}}}

{ 
\rmtwelve

\vsize=9 truein
\hsize=6.5 truein
\raggedbottom
\baselineskip=16pt

February 1994
\vskip2truecm
\centerline
{\titlefont DECOHERENCE PRODUCES COHERENT STATES: }
{\titlefont AN EXPLICIT PROOF FOR HARMONIC CHAINS
\footnote{$^\dagger$}{\rm Published in Phys. Rev. E, 50, 2538 (1994)}} \nobreak
  \vskip 1.0truecm
 
  \author{Max Tegmark}
  \bigskip
  \addr{Max-Planck-Institut f\"ur Physik, F\"ohringer Ring 6}
  \addr{D-80805 M\"unchen;}
  \addr{email: max@physics.berkeley.edu}

  \bigskip
  \centerline{and}
  \bigskip

   \author{Harold S. Shapiro}
  \addr{Department of Mathematics, Royal Institute of
Technology,}
  \addr{S-11044 Stockholm, Sweden;}
  \addr{email: shapiro@math.kth.se}

  \bigskip
 
  \vskip 0.5truecm

\abstract{\rm
We study the behavior of infinite systems of coupled
harmonic oscillators as the time $t\to\infty$, and generalize 
the Central Limit Theorem (CLT) to show that their reduced Wigner
distributions become Gaussian under quite general conditions.
This shows that generalized coherent states tend to be produced naturally. 
A sufficient condition for this to happen is shown to be that the 
spectral function is analytic and nonlinear. 
For a chain of coupled oscillators, the nonlinearity
requirement means that waves must be dispersive, so that 
localized wave-packets become suppressed. 
Virtually all harmonic heat-bath models
in the literature satisfy this constraint, and we have good
reason to believe that coherent states and their generalizations
are not merely a useful analytical tool, but that
nature is indeed full of them.
Standard proofs of the CLT rely heavily on the fact that
probability densities are non-negative. Although the CLT
is generally not applicable if the densities are allowed to take negative
values,  we show that a CLT does indeed hold for a special class of such
functions. We find that, intriguingly, nature has arranged things so
that all Wigner functions belong to this class. 
}
\bigskip
\centerline{PACS Codes: 5.30.-d, 5.30.ch, 2.50.+s, 3.65.-w}

}  
\vfill\eject
 
%
%

\baselineskip12pt
\hoffset0.5cm
\hsize 15.5cm

\beginsection{I. INTRODUCTION}

The phenomenon of decoherence and the useful quantum states known as coherent
states have been extensively studied quite separately, both being interesting in
their own right, and the linguistic similarity of the names may by no more
than a coincidence. Yet it is becoming increasingly clear that the link
between decoherence and coherent states is quite a close one 
\ZurekHabibPaz=\refnr
--- see {\ref} (Zurek, Habib \& Paz 1993, hereafter ZHP) and references therein.
ZHP give
an excellent and up-to-date
discussion of this link, and indicate that decoherence may indeed {\it produce}
coherent state, since it is shown that the
latter tend to be the most
robust states when subjected to interactions with other systems.
This link appears to have been 
first pointed out by K\"ubler and Zeh \ref.
In this paper, we will in a sense
complete this justification of the use of coherent states and their
generalizations, by explicitly proving that they 
are created under quite generic
circumstances.  

\subsection{1.1. Decoherence}

Decoherence refers to some of the changes in a system that are due to
its interaction with its environment. 
Such effects may include suppression of off-diagonal
elements in the spatial density matrix (which makes the system appear more
``classical") and increase in entropy. 
Decoherence is now widely recognized as a key to
the relationship between the quantum and classical realms of physics 
(see {\ref} and references therein). 
Sources of decoherence discussed
in the literature include scattering 
($[\number\refnr-\nextref\nextref\nextref\number\refnr\nextref]$ and others)
and quantum gravity (for instance 
$[\number\refnr,\nextref\number\refnr\nextref]$), but most of the literature has
focused on systems with quadratic Hamiltonians, typically coupled harmonic
oscillators in a chain or some other simple configuration. 
One reason for this
is that systems with quadratic Hamiltonians are just about the only quantum 
systems whose time evolution
can be found analytically. Hence they have provided useful and tractable models.
This is why harmonic chains will be the model of choice in 
the present paper as well. 

Before the interest in decoherence, the main motivation for studying harmonic
chains was the pursuit of a dynamical basis for
equilibrium statistical mechanics. 
An excellent summary of
the early developments in this area is given in
\Huerta=\refnr\ref. A recent summary of subsequent work is given in {\ref}
(Tegmark \& Yeh 1994, hereafter TY), and \FLO=\refnr{\ref} gives a more
comprehensive review. In decoherence applications, the basic calculational
procedure is identical to that in the statistical mechanics applications
mentioned above: The idea is to study the time evolution of some small subset of
the oscillators, called the {\it system}, by taking a
partial trace over the rest of the oscillators, called
the {\it heat bath} or the {\it environment}.
In statistical mechanics applications, the goal is to 
investigate whether the system  
exhibits standard thermodynamic
features such as Brownian motion and approach to thermal
equilibrium.
In decoherence applications, the emphasis is on the behavior of the
reduced density matrix of the system and on the extent to which certain
quantum phase correlations are destroyed.

\subsection{1.2. Generalized coherent states}

For historical reasons, states whose Wigner functions 
$[\number\refnr-$\nextref\HillaryClinton=\refnr\nextref$\number\refnr]$
\KimNoz=\refnr
\nextref
are Gaussian
have been given many different names. The single-oscillator
ground state is a Gaussian centered on the origin. 
When translated in the $q$ and $p$ directions in phase space, 
it is usually
called a coherent state. When rescaled so that it is
shortened in the $q$ direction and elongated in the
$p$ direction (or vice versa), it is known as a squeezed
state. When subjected to the most general linear canonical
transformation (translated, squeezed, and rotated), it is
sometimes known as a tiltedly squeezed state. 
When expanded, it is called a thermal state, and is no longer
pure. The translated ground state of a many-oscillator system is
sometimes called a multimode coherent state. And so on. Thus
the most general state with a Gaussian Wigner function might be
termed a multimode tiltedly squeezed mixed state. We will
simply refer to all these states as 
{\it generalized coherent states}, or 
{\it Gaussian states} for short.

As is indicated by the profusion of names for them, Gaussian states have been
intensely studied in many areas of physics, from quantum optics to statistical
mechanics. 
One reason for this is (just as with harmonic chains) analytic
tractability: if a state is Gaussian at some given time, it will always remain
Gaussian if the Hamiltonian of the system is quadratic, so it is sufficient to
compute the time-evolution of the mean and the covariance matrix, which specify
the Gaussian uniquely. 
Another reason for their popularity is that
coherent states,  
invented by Schr\"odinger in 1926 \ref{} and further developed by 
Glauber \ref, have been seen as a clue to understanding the
classical limit of quantum mechanics. This is because they,
as opposed to for instance energy
eigenstates, exhibit fairly ``classical" behavior.

\subsection{1.3. The connection}

Another Gaussian distribution, the
Maxwell-Boltzmann velocity distribution, is well-known to arise
dynamically from the interactions of many independent
particles, along the lines of the 
Central Limit Theorem. Thus, in the spirit of ZHP, a
natural question to ask is whether generalized coherent states also tend 
to be produced dynamically, from interactions within many-body systems.
In this paper, we will address this question in a case where 
much of the necessary mathematical machinery is already in place: 
the case where the many-body system is a harmonic chain. 

The paper is organized as follows:
In Section II, we review some basic results about
classical and quantum harmonic chains and establish some
notation.
In Section III, we prove the main result of the paper for the
classical case.
In Section IV, we show that the same result is true for the
quantum-mechanical case as well. 
Finally, in Section V, we 
give a more heuristic and qualitative discussion of what
happens for finite systems and for chains lacking
translational invariance.
Some necessary mathematical results are proven in the appendices: 
In Appendix B we place a constraint on the dispersion relationship, 
and in Appendix C
we prove a generalization of the Central Limit Theorem 
for the case where the
``probability density" can take negative values.

\beginsection{II. THE GENERAL HARMONIC CHAIN}

In this section, we establish some
notation and review some basic results about
classical and quantum harmonic chains and cyclic
matrices.

As our quantum system, let us take $2N+1$ coupled harmonic
oscillators of equal mass, labeled 
${-N,...,-1,0,1,...,N}$.
Denoting a point in
the $2\times(2N+1)$-dimensional phase space by 
$$\vz = \pmatrix{\vq\cr\vp}\eq$$ 
and the corresponding operators by
$$\oz = \pmatrix{\oq\cr\op},\eq$$
we can write the Hamiltonian as
\HamEq=\eqnr
$$\opH = {1\over 2m}\op^T\op + 
{m\om^2\over 2}\oq^T A\oq,\eq$$ 
where the time-independent matrix $A$ is symmetric and positive
definite. Throughout this paper, we will use units
where $m=\om=\hbar=1$. The number of oscillators can be
either finite or infinite, but we will limit ourselves to the
infinite case except in Section V.

At any given time, we will specify the (pure or mixed) state of
the system by its Wigner function $W(\vz)$. It is well-known that
since the Hamiltonian is quadratic, the equation of motion
for the Wigner function is identical to that of the
Liouville function in classical statistical mechanics and
has the solution
\EvolEq=\eqnr
$$W_t(\vz) = W_0(U(t)^{-1}\vz),\eq$$
where the time-evolution matrix $U$ is given by\footnote
{$^\dagger$}
{
Here and throughout this paper, the action of a function on a
symmetric matrix is  defined as the corresponding real-valued
function acting on its eigenvalues:
Since all symmetric matrices $A$ can be diagonalized as
$$A=R\Lambda R^T,$$
where $R$ is orthogonal and $\Lambda = \diag{d_i}$ 
is diagonal and real, we can extend any mapping $f$
on the real line to symmetric matrices by defining
\DefEq=\eqnr
$$f(R\diag{d_i}R^T) \equiv
R\diag{f(d_i)}R^T.\eq$$
It is easy to see that this definition is consistent with
power series expansions whenever the latter converge.
For example, 
$$\cos\Aoh = \sum_{n=0}^{\infty}{(-1)^n\over(2n)!}A^n.$$
} 
\Ueq=\eqnr
$$U(t) =
\pmatrix{X&Y\cr
Z&X} \equiv
\pmatrix{\cos\Aoht&A^{-1/2}\sin\Aoht\cr 
-A^{1/2}\sin\Aoht&\cos\Aoht}.\eq$$
By a {\it Gaussian state} in $n$ dimensions
(we will often have $n<2N+1$ further on,
when dealing with reduced Wigner
functions),  we will
mean a state whose Wigner function is Gaussian, {\ie} is of the form
$$W(\vz) = (2\pi)^{-n}(\det C)^{-1/2}
\exp\left[-{1\over 2}(\vz-\vm)^T C^{-1}(\vz-\vm)\right].\eq$$
Here the mean vector $\mu$ and the covariance matrix $C$ 
satisfy
$$\cases{
\vm&$=\expec{\oz}$,\cr
\noalign{\vskip 4pt}
C&$= 
\pmatrix{
\expec{\oq\oq^T}&\oh\expec{\oq\op^T+\op\oq^T},\cr
&\cr
\oh\expec{\oq\op^T+\op\oq^T}&\expec{\op\op^T}
}
 - \vm\vm^T$.}\eq$$
(The symmetric ordering is necessary since $\oq$ and $\op$ do not
commute.)
The Wigner function being Gaussian is equivalent to the density
matrix being Gaussian in the position (or momentum)
representation. 

By a {\it time-independent state}, we will mean
a state with a time-independent Wigner function (or,
equivalently, with a time-independent density matrix). In TY it is shown that 
a necessary but not sufficient condition for a state to be
time-independent is that \SteadyEq=\eqnr
$$\cases{
\vm& = $0$,\crr
C& = $\pmatrix{D&0\cr 0&AD}$,}\eq$$
where $D$ is some constant, symmetric, positive definite matrix
that commutes with $A$. 
If the state is Gaussian, then
this is evidently also a sufficient condition, since the Wigner
function is completely specified by $\vm$ and $C$. 
We will assume that all states have
$\vm= 0$. This in no way reduces the generality of our
treatment, as the time-evolution of $\mu$ and the
time-evolution of the shape of the Wigner function (about
its center $\mu$) are totally independent (see TY). Thus assuming
$\mu=0$ is much like assuming that the center of mass is at rest
at the origin when studying the motion of a blob of jello in the
absence of external forces.

As is conventional, we will assume that the harmonic chain 
is translationally invariant. This is equivalent to the potential matrix
$A$ being {\it cyclic}\footnote
{$^\dagger$}
{Such matrices are often called {\it circulant} in the mathematics 
literature \ref.}, 
{\ie} that each row is a cyclic 
permutation of the row above it: $A_{i+1,j+1} = A_{ij}$, understood 
(mod $n$) for an $n\times n$ matrix.
Since $A$ is also symmetric, this
means that we can
write $A_{ij} = a_{|i-j|}$ and interpret the system as a chain of
harmonic oscillators where the coupling between any two
oscillators depends only on the separation between them. 
(If $N$ is finite, we can interpret the system as oscillators arranged in a
ring rather than a line.)
Using 
$(5)$, we can write any function of a
(cyclic or non-cyclic) matrix $A$ as 
\fmnEq=\eqnr
$$f(A)_{mn} = \sum_k R_{mk}R_{nk}f(d_k).\eq$$
Cyclic matrices have the great
advantage that they all commute. This is because 
they can all be diagonalized by the same matrix $R$, an
orthogonal version of the discrete Fourier matrix. 
Physically, this means that plane waves form a complete set of
solutions.
If $A$ is
symmetric, positive-definite, cyclic and infinite-dimensional,
then Eq. $(\number\fmnEq)$ reduces to \FKM=\refnr\ref 
\FKMeq=\eqnr
$$f(A)_{mn} = \ootp\ipp d\theta
f\left[\lth\right]
\cos(m-n)\theta,\eq$$ 
where the spectral function $\lth$ is 
the function whose Fourier coefficients are row zero of $A$.
The spectral function can be interpreted as
a dispersion relationship, $\lambda$ being the frequency of a
wave with wave number $\theta$. 
Note that $f(A)$ is cyclic as well, {\ie} its
components depend only on the distance to the diagonal.

A cyclic potential frequently discussed in the literature is
the {\it nearest neighbor potential}, the 
case where each mass is coupled only to a fixed spring and to
its nearest neighbor: 
$$\opH = \sum_{k=-\infty}^{\infty} \left[{1\over 2}\hat p_k^2 + 
{1\over 2} \hat q_k^2 +
{\gamma^2\over 2}
\left(\hat q_{i+1}-\hat q_{i}\right)^2\right],\eq$$  
{\ie} $A_{kk} =
1+2\gamma^2$, $A_{k,k\pm1} = -\gamma^2$ and all other elements
of $A$ vanish. For this special case, the spectral function is  
$$\lambda^2(\theta) = 1 + 4\gamma^2\sin^2{\theta\over
2}.\eq$$

\beginsection{III. THE INFINITE CLASSICAL CHAIN}

In this section, we will investigate the circumstances 
under which states become Gaussian in classical statistical
mechanics. Here the positions and momenta at time $t$ are
specified by $\vz(t)$, which is a vector of random
variables.
These random variables are given by the initial random
variables as  
$$\vz(t) = U(t)\vz(0),$$
and we wish to study the circumstances under which 
the probability distribution of $\vz(t)$ becomes a multivariate
Gaussian as $t\to\infty$.

According to equation $(\number\Ueq)$,
the position of oscillator $m$ at time $t$
is given by the initial data as  
\xiEq=\eqnr
$$q_m(t) = \sumnoi
\xi_{mn}(t),\eq$$ 
where we have defined the random variables
$$\xi_{mn}(t) \equiv X_{mn} q_n(0) + Y_{mn} p_n(0).\eq$$
(The above expression is to be understood without any
summation.)
Using the Liapunov version of the  
Central Limit Theorem
\Renyi=\refnr
({\it e.g.} {\ref}), we see that the distribution
of $q_m(t)$ becomes Gaussian as $t\to\infty$ if the Liapunov
condition 
$${M^{(3)}\over M^{(2)}} \to 0\as t\to\infty\eq$$
is satisfied, where we have defined
$$M^{(k)}(t)\equiv
\left[\sum_{n=-\infty}^{\infty}\expec{|\xi_{mn}(t)|^k}\right]^{1/k}.\eq$$ 
If we make the
physically reasonable assumption about the second and
third moments of the initial data that $\expec{z_k(0)^2}^{1/2} >
\sigma$  and $\expec{|z_k(0)|^3}^{1/3} < \kappa$ for some
positive constants $\sigma$ and $\kappa$, then then the
Liapunov condition reduces to the requirement that
\LiapunovEq=\eqnr
$${
\left(\sumnoi 
\left|X_{mn}\right|^3 + 
\left|Y_{mn}\right|^3\right)^{1/3}
\over 
\left(\sumnoi 
X_{mn}^2 + 
Y_{mn}^2\right)^{1/2}
}
\to 0\as t\to\infty.\eq$$
(For the quantum case to be treated in the next section, the
assumption of a minimum standard deviation follows directly from
the Heisenberg uncertainty principle, if we simply assume that
the standard deviations are bounded from above.) 

If $A$ is cyclic, then $(\number\Ueq)$ and 
$(\number\FKMeq)$ yield
$$\cases{
X_{mn}& = $\ootp\ipp
\cos[\lambda(\theta)t]
\cos(m-n)\theta\>d\theta$,\crr
Y_{mn}& = $\ootp\ipp
\lambda(\theta)^{-1}\sin[\lambda(\theta)t]
\cos(m-n)\theta\>d\theta$.
}\eq$$
Now Parseval's theorem gives 
$$\sumnoi 
X_{mn}^2 + Y^2_{mn} = 
\ootp\ipp
\left(\cos^2[\lambda(\theta)t] +
\lambda(\theta)^{-2}\sin^2[\lambda(\theta)t] \right)
\>d\theta,\eq$$
which approaches some positive constant $c$ as $t\to\infty$.
Thus the Liapunov condition is satisfied if and only if the
numerator of Eq. $(\number\LiapunovEq)$ approaches zero as
$t\to\infty$.  But 
$$\sumnoi\left|X_{mn}\right|^3 + \left|Y_{mn}\right|^3 \leq 
\left(\sup_n \left|X_{mn}\right| + \left|Y_{mn}\right|\right)
\sumnoi 
X_{mn}^2 + Y^2_{mn},\eq$$
so it suffices to show that this supremum approaches
zero\footnote{$^\dagger$} {This is not merely a sufficient
condition but also a necessary condition for $q_m(t)$ to become
Gaussian, since a non-zero supremum means that there is some 
$\xi_{mn}$ that makes a finite contribution to the sum
$(\number\xiEq)$. This would imply that the distribution of
$q_m(t)$ depends on the details of the distribution of
$\xi_{mn}$ and thus cannot in general be Gaussian.}.
In Appendix A we show that this supremum does
indeed approach zero under quite general conditions, namely for
any spectral function $\lambda$ that is analytic on the
entire interval $[-\pi,\pi]$ and in addition is non-linear. 

In conclusion, we have shown that the probability
distribution of $q_m(t)$ becomes Gaussian as $t\to\infty$ if
the spectral function $\lambda$ is non-linear and analytic on
$[-\pi,\pi]$ and if the initial probability distributions of
all positions and momenta are independent and have bounded
second and third moments. 
The proof that $p_m(t)$ becomes Gaussian is completely
analogous. 

The assumption that all random variables are
independent can be relaxed 
to assuming
that no dependence exists at oscillator separations larger than
some fixed integer $M$. More precisely
$[\number\refnr,\nextref\number\refnr\nextref]$, it is sufficient that
there exists an $M$ such that $n-m > M$ implies that the two
infinite sets
$$(...,q(0)_{m-1},p(0)_{m-1},q(0)_m,p(0)_m)
\quad\hbox{and}\quad
(q(0)_n,p(0)_n,q(0)_{n+1},p(0)_{n+1},...)$$
are independent. Finally, our proof can readily be generalized 
by using the multivariate CLT 
$[\number\refnr,\nextref\number\refnr\nextref]$
to show that all finite multivariate distributions become Gaussian.

\beginsection{IV. THE INFINITE QUANTUM CHAIN}

In this section we will see that all the results of
the previous section can be generalized to the
quantum-mechanical case. Much of the mathematics remains
the same, but the interpretation changes. The big mathematical
difference is that a Wigner function can take negative
values, whereas a classical probability distribution
cannot. A generalization of 
the Central Limit Theorem for Wigner distributions is 
proved in Appendix C. 

By analogy with reduced density matrices, all expectation values
of the $n^{th}$ oscillator can be calculated from the $n^{th}$
single oscillator {\it reduced Wigner function} 
$[\number\HillaryClinton,\number\KimNoz]$
\ReductionEq=\eqnr
$$W^{(n)}(q_n,p_n) \equiv \int_{(n)}W(\vq,\vp),\eq$$
where the integral is to be taken over all variables 
{\it except} $x_n$ and $p_n$.
This is analogous to the way the marginal probability
distribution for $(x_n,p_n)$ is calculated in classical
statistical mechanics. The only difference is that the Wigner
function can take negative values and cannot be interpreted as
a probability distribution. 
In Section III, we gave necessary
and sufficient conditions for when various marginal
distributions become Gaussian as $t\to\infty$. Here
we will pursue the quantum analog and give conditions for when
various reduced Wigner functions become Gaussian.

Fourier transforming equation $(\number\ReductionEq)$
with respect to all variables yields
\FourierReductionEq=\eqnr
$$\Wh^{(n)}(q_n,p_n) = 
\widehat{W}(0,...,0,q_n,0,...,0,p_n,0,...,0),\eq$$
{\ie} the Fourier transformed Wigner function (also known as
the {\it characteristic function}) with all variables except 
$q_n$ and $p_n$ set equal to zero. This
expression is often more useful than 
$(\number\ReductionEq)$, as it contains no
integrals. Fourier transforming equation $(\number\EvolEq)$ and
using the fact that det $U$ = 1 yields \FourierEvolEq=\eqnr
$$\Wh_t(\vz) = \Wh_0\left[U(t)^T\vz\right],\eq$$ 
where $U^T$ denotes the
transpose of $U$. 

Let us first assume that the oscillators are not entangled 
initially, so that the Wigner
function for the initial state is completely separable, {\ie} of the form
\SepEq=\eqnr
$$W_0(\vq,\vp) = \prod_n W_0^{(n)}(q_n,p_n)\eq$$
for some set of reduced Wigner functions $W_0^{(n)}$. 
Substituting equation  $(\number\SepEq)$ into equations
$(\number\FourierReductionEq)$ and $(\number\FourierEvolEq)$
yields
$$\Wh^{(m)}_t(q_m,p_m) =  
\prod_{n} \Wh_0^{(n)}(X_{mn}q_m + Z_{mn}p_m, Y_{mn}p_m + X_{mn}p_m),\eq$$
where the matrices $X$, $Y$ and $Z$ are those
defined in equation $(\number\Ueq)$, and no summation is implied.
Thus the reduced Wigner function is
obtained by Fourier transforming the initial reduced Wigner
functions, multiplying them together, rescaling their
arguments appropriately, and performing an inverse Fourier
transform on the result. This is exactly how we would compute
the probability density for a weighted sum of independent
two-dimensional random variables, which is the classical case that we
investigated in the previous section.
The standard versions of the CLT all make heavy use of the assumption that
probability densities are non-negative. 
Thus in order to show that the reduced Wigner
function becomes Gaussian, we need a Liapunov type CLT
for ``random variables" whose ``probability densities"
are allowed negative values, a subject which to our knowledge has not
been previously studied. We leave the full mathematical
details of such a
study for a future paper, but prove such a generalized CLT 
in Appendix C for the special case where 
all the ``random variables" are identically distributed. 
It appears highly plausible that the 
standard Liapunov proof can be appropriately generalized employing
similar techniques.

In conclusion, this would show that any one-particle 
reduced Wigner function $\Wh^{(n)}_t(q_n,p_n)$ 
becomes Gaussian as $t\to\infty$ if
the spectral function $\lambda$ is non-linear and analytic on
$[-\pi,\pi]$ and if the initial states of
all oscillators satisfy the condition that certain
expectation values be bounded as described in Section III.
Specifically, the expectation values of all linear, quadratic and cubic
combinations of $\widehat{q}_n$ and $\widehat{p}_n$ should be 
bounded from above
by some constant independent of $n$. 
Then as shown in Theorem III in appendix C,
the other moment constraints will be automatically satisfied because of the
Heisenberg uncertainty relationship.

Just as in the classical case,  
the assumption that no initial correlations exist between
different oscillators can be relaxed to
assuming that the joint Wigner functions are separable for
oscillator separations larger than some fixed integer $M$.
The generalization to the reduced Wigner function for more than
one particle is also completely analogous.

Our result shows that virtually all harmonic chains treated in
the literature will produce Gaussian states as $t\to\infty$,
since they tend to have spectral functions that are both
analytic and nonlinear. Some well-known examples of such
harmonic chains are the above-mentioned nearest neighbor model 
$[\number\refnr\nextref,\number\Huerta]$
and the FKM model $[\number\FKM]$.
Since the FKM model has
been shown to be equivalent to the independent-oscillator heat
bath model $[\number\FLO]$, the latter will also produce Gaussian states
under quite general conditions. 

An interesting mathematical problem is to generalize 
our results to arbitrary quadratic
systems, by giving conditions for when they produce Gaussian
states. It is our belief that Gaussian states
will be seen to be produced under quite generic
circumstances, and thus are ubiquitous whenever there is
interaction between a very large number of systems.

In TY, it is shown that 
if a harmonic chain starts out with an arbitrary cyclic 
covariance matrix
$$C = \pmatrix{E&G\cr G&F},$$
then 
$$C \to
\pmatrix{D&0\cr 0&A D}
\quad\hbox{as }t\to\infty,\eq$$
where
$$D = \oh\left[E+ A^{-1}F\right].$$
If the spectral function is nonlinear and analytic as discussed
above, the convergence will not merely be
pointwise as shown in the TY, but indeed
uniform.
Since a Gaussian is uniquely specified by its mean vector 
$\vm$
and its covariance matrix $C$, we thus know not only that
the harmonic chain approaches a Gaussian state, but also exactly
which Gaussian state. As we would expect, the only information
that is preserved about the initial data is the second moments,
{\ie} the covariance matrix, whereas all fine details of the
Wigner function and all information about higher moments are
lost. Note that the initial data enter only in the combination
$E+A^{-1}F$, so all information about $G$ (initial 
position-momentum correlations) 
is lost as well.

Without loss of generality, we assumed that the mean vector
$\vm=0$ in the above treatment. 
The effect of relaxing this assumption is discussed in TY. It is
seen that whereas the covariance matrix still converges to the
value given above (and from what we have shown, all higher
central moments converge to the values required by 
Gaussianity), the mean vector $\mu$ does
{\it not} converge towards a constant, but keeps oscillating
forever.

\beginsection{V. DISCUSSION}

In this paper, we have shown that any part of a generic harmonic
chain will evolve into a Gaussian state as $t\to\infty$. 
Given that the spectral function is mathematically well-behaved
(analytic on the interval $[-\pi,\pi]$), ``generic" is to be interpreted
as forbidding two special cases:

1) The spectral function is linear.

2) Fine-tuned long-range correlations exist in the initial data. 

\noindent
We will now attempt to give a more intuitive and physical interpretation 
of these two conditions (which apply for infinite chains), 
as well as qualitatively discuss what happens if $N$ is
large but finite.

The gist of the CLT as we have used it is that a weighted
average of infinitely many independent
random variables approaches Gaussianity as 
$t\to\infty$ if all weights become infinitesimal. 
Very loosely speaking, a sum of infinitely
many infinitesimally small independent random contributions is Gaussian. 
In terms of our harmonic chains, information
about the initial data must be mixed, and mixed so thoroughly that the state of
any subsystem of the chain at $t=0$ will have only an infinitesimal impact on the
state of any subsystem of the chain as $t\to\infty$. 
Physically, what can go wrong? In the extreme case $A\propto I$, 
which corresponds to the oscillators being completely uncoupled, 
there is no mixing of information whatsoever and the CLT fails miserably. 
Now one might think that as long as an oscillator is 
coupled to at least one other
oscillator (and thus indirectly to an infinite number of oscillators through
it, by translational invariance), the CLT should always apply, and Gaussians
should be obtained for any cyclic potential matrix {\it except} $A\propto I$. 
This is false.
If the spectral function is linear (or, dropping the analyticity requirement, if
it is linear on any finite interval), then a wave-packet composed
only of wavenumbers in this interval will simply travel down the chain without
dispersing, retaining its initial shape forever. Thus the initial data at one
point will have a non-infinitesimal impact on the state somewhere else, even at
arbitrarily late times. 
This is reflected as $U_{mn}\to 0$ as $t\to\infty$ for any fixed $m$ and $n$
as shown in TY, while 
$\sup_{m,n} |U_{mn}|$ remains bounded away from zero, 
as elements of order unity
merely propagate further and further away from the diagonal, at a linear rate. 
In summary, the key is that the propagation of waves must be 
{\it dispersive}, {\ie} the dispersion relationship must be non-linear. 
This will ensure that all localized wave packets gradually get destroyed. 

The second constraint, that on the initial data, is closely related to 
the second law of thermodynamics: although for most initial data, the
entropy of isolated gas in a container will not decrease, 
there is a small set
of rather contrived initial data for which it will, and time will appear to run
backwards for a while. The easiest way to obtain such initial data is to let a
low-entropy state evolve into a high-entropy state and then reverse all
velocities. The situation with our harmonic chains is completely analogous: If
an uncorrelated state is allowed to evolve, the entropy of the
subsystems will increase as each oscillator becomes increasingly correlated
with ever more distant neighbors. If we now replace 
$W(\vq,\vp)$ by exactly $W(\vq,-\vp)$ (approximately will not suffice), 
the system will evolve back into the uncorrelated (and perhaps non-Gaussian)
system we started with. Apparent time-reversal is always caused by such
long-range correlations, and since
we used a version of the CLT that bans such correlations, 
such troubles are avoided altogether. 
Of course, after the uncorrelated initial state has been obtained, 
new correlations begin to arise again, and the subsystems eventually approach
Gaussianity. An interesting problem is to investigate whether, in this vein, 
our result can be proven to hold for any cyclic initial conditions
whatsoever.

The result that subsystems become Gaussian as $t\to\infty$ holds strictly only
for infinite chains. So what happens when $N$ is finite but very large? 
If the waves are dispersive, then the discussion of finite $N$ in 
TY can readily be extended to show that
$\max_{m,n} U_{mn}$ will evolve as follows when $N$ is large:

\llist{
(i) During an initial transition period whose duration
is of the order of the dynamical time scale $\omega_0^{-1}$, it
decays from its initial value of order unity to a value of order
$N^{-1/2}$.}

\llist{
(ii) After that, it oscillates around this value with an
oscillation amplitude of the same order.}

\llist{
(iii) Since the time evolution of $U_{mn}$ is almost periodic, 
some components must 
return to values of order unity an infinite number of times.
This happens approximately once every Poincar\'e recurrence
time.
However, as shown by {\ref}, the Poincar\'e 
time scale is generally enormous compared to the dynamical
time scale, since it tends to grow exponentially with $N$ for
systems of this type.
}

In a discussion of density matrices {\ref}, Feynman writes:
``When we solve a
quantum-mechanical problem, what we really do is
divide the universe into two parts -- the system in which we are
interested and the rest of the universe. We then usually act as if
the system in which we are interested comprised the entire
universe."
In this spirit we summarize our harmonic chain result:
The effect of ``the rest of the universe" is to make our subsystem
approach a generalized coherent state. 
Since most
systems in the real world are coupled to their
environment, this gives us even more reason to believe
that nature is indeed full of generalized coherent states.

\bigskip
The authors would like to thank Emory Bunn, Leehwa Yeh and 
Wojtek Zurek for useful comments on the manuscript.

\beginsection{APPENDIX A}

\def\R{{\bf R}}
\def\C{{\bf C}}
\def\as{\quad\hbox{as}\quad}
\def\Ca{C_1}
\def\Cb{C_2}
\def\Cc{C_3}
\def\Cd{C_4}
\def\Ce{C_5}
\def\Cf{C_6}
\def\Cg{C_7}
\def\Ch{C_8}
\def\l{\ell}
\def\e{\varepsilon}
\def\urf{\varphi}

In this appendix we give a condition for when 
$\sup_{m,n} \left|U(t)_{mn}\right|\to 0$
as $t\to\infty$. This rests on Theorem (I),
which is proved in Appendix B.
Since according to Eq. $(\number\Ueq)$ and Eq. $(\number\FKMeq)$,
$$\cases{
X_{mn}&= $\ootp\ipp
\cos[\lambda(\theta)t]
\cos[(m-n)\theta]
\>d\theta,$\crr
Y_{mn}&= $\ootp\ipp
\sin[\lambda(\theta)t]
\cos[(m-n)\theta]
\lambda(\theta)^{-1}
\>d\theta,$\crr
Z_{mn}&= $\ootp\ipp
\sin[\lambda(\theta)t]
\cos[(m-n)\theta]
\lambda(\theta)
\>d\theta,$
}$$
we wish to show that 
$$\sup_k\left|
\ipp e^{i\lambda(\theta)t}e^{-ik\theta} g(\theta) d\theta\right| \to 0
\as t\to\infty,$$
where 
$k = \pm(m-n)$ is any integer,  
$g(\theta) = \lambda(\theta)^{\nu}$,
and
$\nu=0$, $\nu=-1$ and $\nu=1$, respectively.
Setting $f(\theta) = \lambda(\theta)$, Theorem I shows that $\sup_{m,n}
\left|X(t)_{mn}\right|\to 0$ as $t\to\infty$ if $\lambda$ is a non-linear analytic
function on the entire interval $[-\pi,\pi]$. 
The same holds for $Z(t)_{mn}$, {\ie} the $\nu=1$ case. Since $A$ is positive
definite, $\lambda$ is bounded from below by some positive constant, so
$\lambda^{-1}$ is also analytic and $\sup_{m,n}
\left|Y(t)_{mn}\right|\to 0$ as $t\to\infty$ follows under the
same conditions.
In summary,
$\sup_{m,n} \left|U(t)_{mn}\right|\to 0$
as $t\to\infty$ for any bounded non-linear analytic spectral
function $\lambda$. 

It is noteworthy that the non-linearity requirement is crucial to
ensure that the convergence to zero is uniform, independent of
$m$ and $n$.  By simply changing variables and using
Riemann-Lebesgue's Lemma, it is readily seen that
$U(t)_{mn}$
will approach zero as $t\to\infty$ for any fixed $m$ and $n$,
even if $\lambda$ is linear. However, as was discussed in
Section V, this alone is not sufficient for producing Gaussian states.

\beginsection{APPENDIX B}

In this Appendix, which is purely mathematical, we prove the basic convergence
theorem upon which the conclusion of the paper rests.
The theorem is of course a version of ``Van der Corput's Lemma"
\Zygmund=\refnr
{\ref}
in the theory of oscillatory integrals, but the uniformity with
respect to $k$ (which we believe is new) requires quite delicate
handling.

{\bf Theorem I.} Let $f$ be a function analytic
on a neighborhood of the closed bounded interval $I$ of the real axis, and
real-valued on $I$. Assume $f$ is not a polynomial of degree $\leq 1$. Then, for
any $g\in C^1(I)$, we have
$$\sup_{k\in\R}\left|
\int_I e^{if(x)t}e^{-ikx} g(x) dx\right| \to 0
\as t\to\pm\infty.\eqno(B1)$$

Actually, as the proof will show, the left-hand side is
$O\left(|t|^{-\sigma}\right)$ for some $\sigma>0$.

In the proof, we may restrict attention to $t>0$, as the other case
then follows if we replace $f$ by $-f$.
By hypothesis, there is an open simply connected domain 
$D$ containing $I$
such that $f$ is analytic on a neighborhood of the closure $\bar D$
of $D$.

{\bf Lemma B1.} There is an integer $\l$ such that, for every $w\in\C$,
$f'(z) - w = 0$ has at most $\l$ roots (counting multiplicities) in
$\bar D$. 

{\bf Proof.} This is a simple exercise in complex analysis.

{\bf Lemma B2.} If $F$ is of class $C^2(J)$ and real-valued for
some closed bounded interval $J\subset\R$, $g\in C^1(J)$, and
$F$ is strictly monotone on $J$, then for $t>0$ we have 
$$\left|\int_J e^{iF(x)t} g(x) dx\right| \leq {\Ca\over\delta^2 t},
\eqno(B2)$$
where $\delta$ denotes the smaller of $1$ and 
$\min|F'(x)|$ for $x\in J$. Here $\Ca$ is a constant depending only
on $g$ and the number $\max |F''(x)| : x\in J$.

{\bf Proof.}
This is a standard estimate of ``Van der Corput type" (see for
instance $[\number\Zygmund]$.
This is a rather primitive version, the proof being a straightforward
variable change $y=F(x)$ followed by partial integration. With stronger
hypotheses one can get $\delta$ rather than $\delta^2$ in the denominator, but
this is not required for our purposes.

In the following, a number of constants whose precise values are
not essential will arise. Constants denoted $C_1$, $C_2$,... will all
be independent of $k$ (later $u$), depending only on the functions
$f$ and $g$ and the geometric entities $I$, $D$.

{\bf Proof of Theorem.}
We must estimate the integral 
$$\int_I e^{if(x)t} e^{-ikx} g(x) dx 
= \int_I e^{iF(x) t} g(x) dx,$$
where  $F(x) = f(x) - ux$, and $u$ = $k/t$ is a real parameter. By
Lemma B1, the number of complex zeroes $z$ to $f'(z)-u = 0$ in $\bar D$
is bounded by an integer $\l$ independent of $u$. Denote the distinct
zeroes by $z_j = z_j(u)$; $j=1,...,s$, with corresponding
multiplicities $m_1,...,m_s$ and $\sum_{j=1}^s m_j \leq\l$. Now, fix
$\e>0$ and let $\Delta_j$ denote an open disk of radius $\e$ centered
at $z_j$. Then, $I\setminus \bigcup_{j=1}^s \Delta_j$ consists of a union 
of $r\leq l+1$ pairwise disjoint closed intervals $J_i$,
on each of which $F$ is strictly monotone. Moreover we have the
estimate 
$$|F'(x)|\geq\Cb\e^{\l}\eqno(B3)$$
for all $x$ in these intervals. We will show this after completion of
the argument. By (B2) we have for small $\e$
$$\left|\int_{J_i} e^{iF(x)t}g(x)dx\right|
\leq {\Cc\over\e^{2\l}t}.$$
Summing over $i$, and noting that $I\setminus\bigcup\,J_i$
has length $\leq 2\l\e$, we get
$$\left|\int_I e^{iF(x)t}g(x)dx\right|
\leq\Cd\left[\e + \left(\e^{2l}t\right)^{-1}\right].\eqno(B4)$$
For fixed (large) $t$, choose here $\e=t^{-1/(2\l+1)}$ and we see
that the left-hand term in (B4) is bounded by 
$\Ce t^{-1/(2\l+1)}$. This concludes the proof of the Theorem. 

We now supply the proof for the estimate (B3). Let us
define the polynomial 
$$P(z)=P(z,u)\equiv\prod_{j=1}^s[z-z_j(u)]^{m_j}.$$
It is clear that there is some constant $\Cf$ such that 
$P(z,u) < \Cf$ for all $z\in\bar D$ and for all $u$. 
Now, consider the function $P(z,u)/(f'(x)-u)$. It is analytic in
$\bar D$. Moreover, for some constant $\Cg$, 
$$\max_{x\in I} {|P(x,u)|\over|f'(x)-u|}\leq\Cg. \eqno(B5)$$
(We will return to the proof of (B5) shortly.) 
Thus, for $x\in I$, 
$$|f'(x)-u|\geq \Cg^{-1} |P(x,u)| \geq \Cb \e^{\l}$$
for some constant $\Cb$
when $x\in I\setminus (U\Delta_i)$. Thus all that remains in order to
prove (B3) is to show that (B5) holds. This can be done as
follows. Let $\Gamma_1$, $\Gamma_2$,...,$\Gamma_{\l+1}$ be
pairwise disjoint simple closed curves in $D$, each of which
encloses $I$.
 
{\bf Lemma B3.}
There is a positive constant $\Ch$ such that for any
$u$, 
$$\min_{z\in\Gamma_j} |f'(z)-u| \geq \Ch$$
holds for at least one value of $j$. 

{\bf Proof. } Let us define $\urf_j(u) \equiv\min |f'(z)-u| :
z\in\Gamma_j$. It is easy to see that $\urf_j$ is continuous. Hence,
so is 
$$\urf(u) \equiv\max_{1\leq j\leq\l+1} \urf_j(u).$$ 
Moreover, $\urf(u) > 0$, because if $\urf(u) = 0$ for some $u$, then
all $\urf_j(u)$ are zero, so $f'(z) - u$ vanishes at least once on
each $\Gamma_j$ and thus has at least $\l+1$ zeroes, a contradiction. 
Since $\urf(u)$ is continuous, positive and obviously $\to\infty$ as
$|u|\to\infty$, it attains a positive minimum value $\Ch$. We thus
have that for every $u$, there is at least one $j=j(u)$ such that 
$\urf_j(u)\leq\Ch$, which proves the Lemma. 

By the maximum modulus theorem, for any $u\in\C$,
$$\max_{z\in I} {|P(z,u)|\over|f'(z)-u|}
\leq \max_{z\in\Gamma_j} {|P(z,u)|\over|f'(z)-u|},
$$
where we choose $j=j(u)$ as in Lemma B3. Thus on the right hand
side, the numerator is bounded from above by $\Cf$ and the 
denominator is bounded from below by $\Ch$, so the entire expression
is $\leq\Cg\equiv \Cf/\Ch$. This completes the proof.

\beginsection{APPENDIX C}

In this Appendix, we prove a generalized version of the Central Limit Theorem
(CLT) that holds for Wigner Distributions. 
Although the CLT can {\it not} be generalized to arbitrary functions that are
allowed to take negative values, we show that repeated convolutions does
indeed lead to Gaussianity for a special class of such
functions. We find that, for some reason, nature has arranged 
things so that all Wigner functions belong to this class. 

Given a function $f$ on $\R^d$, its zeroth, first and second moments
are defined as  
$$\eqalign{
M^{(0)}&\equiv\int f(x)d^dx,\crr
M^{(1)}_i&\equiv\int f(x)x_id^dx,\crr
M^{(2)}_{ij}&\equiv \int f(x)x_i x_j d^dx
}$$
if the moments exist, {\ie} if these integrals are convergent in the Lebesgue
sense. 
In probability theory, the second central moment matrix 
$V_{ij}\equiv M^{(2)}_{ij} - M^{(1)}_i M^{(1)}_j$ is usually called
the {\it covariance matrix}.
Let us define a {\it quasi-probability density} on $\R^d$ as a
real-valued function $f$ having the following
properties: 

\slist{* $M^{(0)} = 1.$}

\slist{* The first moments $M^{(1)}_i$ exist.}

\slist{* The second moments exist and the covariance matrix is strictly
positive definite.}

\slist{* For reasons that will become clear later, we will also make the
technical assumption that $f$ is an $L^2$ function, {\ie} square-integrable.}

\noindent
If $f$ has the additional property that it is non-negative,
{\ie} that it is a probability density, then the basic
version of the CLT states that if we define $f_n$ to be $f$ convolved with
itself $n$ times and translated and rescaled so as to have the same first and
second moments as $f$, then $f_n$ approaches a Gaussian $g$
as $n\to\infty$. 
The convergence is usually shown to be in the 
weak topology of
measures, which in our context means that integral of $f_n$ 
times any bounded testfunction tends to the corresponding 
integral for $g$.
We wish to investigate under which circumstances $f_n$ approaches a
Gaussian if we drop the assumption of non-negativity.

Without loss of generality, we may assume that $M^{(1)}_i = 0$ and that
$V_{ij} = \delta_{ij}$, the identity matrix, as 
the general case can be obtained from this by a simple change of variables.
By Fourier transforming and using the convolution theorem, 
one then obtains the standard expression
\FourierPowerEq=\eqnr
$$\fh_n(\vk) = \fh\left(n^{-1/2} \vk\right)^n.\eqno(C1)$$
Our problem decomposes into two parts:

\llist{(A)
To give conditions for when 
$\fh_n(\vk) \to \gh(\vk) = e^{-k^2/2}$ as $n\to\infty$.
}

\llist{(B)  
To show that this convergence to Gaussianity on the Fourier side
really implies that $f_n \to g$ in some meaningful sense.
}

\noindent
It is important to note that (B) is not merely an unphysical mathematical 
detail. This is illustrated by the following counterexample:
Take $d=1$ and chose $\fh(k)$ to be any smooth, symmetric $L^2$
function 
such that $\fh(0)=1$, $\fh'(0)=0$,
$\fh''(0)=-1$ and $\fh(k_*)>1$ for some constant $k_*>0$. 
An example of such a function is $\fh(k) = (1+k^4)e^{-k^2/2}$.
It is easy to see that its
inverse Fourier transform $f$ will have all the properties of a 
quasi-probability density. 
It is also easy to show that  $f_n(k)\to g(k)$ as
$n\to\infty$ pointwise, for any fixed $k$, since $\fh(k) = 1 - k^2/2 + O(k^3)$
follows from our assumptions, and 
$$\left(1 - {k^2\over 2n}\right)^n\to
e^{-k^2/2}\as n\to\infty.$$ 
Yet Eq. $(C1)$ clearly shows that
the part of the curve that exceeds unity will grow ever larger as $n$ increases. 
Pointwise
convergence is obtained merely because the growing $|\fh_n|>1$ hump keeps
shifting out to higher and higher frequencies $k$. Thus as $n$ grows large,
$\fh_n$ may look quite Gaussian on the interval $|k|\ll n^{1/2} k_*$,
but there will be exponentially growing bumps of height $\fh(k_*)^n$ 
at  $k = \pm n^{1/2} k_*$. Inverse Fourier transforming, this means
$f_n$ will behave like a sum of a Gaussian and violent noise, whose
frequency and amplitude increase without bounds as $n\to\infty$. 

We will refer to a quasi-probability density as {\it proper}
if the absolute value of its Fourier transform takes its maximum only at the
origin.  Thus $f$ is proper if $|\fh(\vk)|\leq 1$, with equality only for $\vk=0$.
If a quasi-probability density never takes negative values (and hence is a
probability density in the conventional sense), then it is easy to show that
it will automatically be proper.  
The ``ultraviolet catastrophe" described above shows that a
necessary condition for a CLT to hold is that 
$|\fh|$ never exceeds unity.
Thus being proper is a necessary condition, except perhaps for the
borderline case where $|\fh(\vk)|\leq 1$ but actually equals unity for some 
$\vk\neq 1$.
In what follows, we will show that being proper is also a
sufficient condition. We will
also see that, interestingly, all Wigner quasi-probability densities are
proper. 

In what follows, the function $g$ will always denote the $d$-dimensional Gaussian 
$$g(\vx) \equiv {1\over(2\pi)^{d/2}} e^{-x^2/2}.$$
Unless otherwise indicated, all integrals below are to be taken over all space. 
$||\cdot||_2$ will denote the $L^2$ norm in $\R^d$, defined by 
$$||f||_2 \equiv \left(\int|f(\vx)|^2 d^dx\right)^{1/2}.$$

{\bf Lemma C1:}
If $f$ is a proper quasi-probability density, then for any
$\varepsilon>0$, there exists a $\delta>0$ such that $|\fh(\vk)| \leq 1-\delta$
for all $|\vk|>\varepsilon$. 

{\bf Proof:}
If $f$ is integrable, then $\fh(\vk) \to 0$ as $|\vk|\to\infty$ by
Riemann-Lebesgue's Lemma, so the continuous function $|\fh(\vk)|$ attains some
maximum value $M_{\varepsilon}$ on the set 
$\lbrace k:|\vk|\ge\varepsilon\rbrace$.  $M_{\varepsilon} < 1$ since
$f$ is proper, so we can choose $\delta = 1-M_{\varepsilon}$.
Alternatively, if we do not wish to assume that $f$ is integrable,
it is straightforward to show that $\fh(\vk) \to 0$ as $|\vk|\to\infty$ 
if $f$ is any Wigner function. 

{\bf Lemma C2:}
The norms $||\fh_n||_2$ are bounded by a constant independent of $n$.

{\bf Proof:}
For small $\vk$, $\fh$ has the asymptotic behavior 
$\fh(\vk) = 1 - k^2/2 + o(k^3).$ 
Thus it is easy to see that given any
constant $p<1$, there exists an $\varepsilon_p$ such that 
$$|\fh(\vk)|^2\leq e^{-pk^2}$$
for all $|\vk| \leq \varepsilon_p$. For all other $\vk$, we have
$$|\fh(\vk)|^2 \le (1-\delta_p)^2$$ for some $\delta_p>0$ by Lemma C1.
Combining these two bounds, we obtain
$$||\fh_n||_2^2 = \int |\fh_n(\vk)|^2\dk \le
\int_{|\vk|\le\sqrt{n}\varepsilon_p} e^{-pk^2}\dk
+ (1-\delta_p)^{2(n-1)}
\int_{|\vk|>\sqrt{n}\varepsilon_p}|\fh(n^{-1/2}\vk)|^2\dk.$$ 
Extending both
integrals to all of space and changing variables in the second one, we get
\NormBoundEq=\eqnr
$$||\fh_n||_2^2 \leq 
||e^{-pk^2/2}||_2^2 + 
n^{d/2}(1-\delta_p)^{2(n-1)}\>||\fh||_2^2.\eqno(C2)$$
Since the last term $\to 0$ as $n\to\infty$, the left hand side is bounded
by a constant independent of $n$.

{\bf Lemma C3:}
$\fh_n(\vk)\to \gh(\vk)$ pointwise as $n\to\infty$.

{\bf Proof:}
This step is identical to that in proofs of the classical CLT (see for instance 
$[\number\Renyi]$), so we omit it.

{\bf Lemma C4:}
$\fh_n\to\gh$ in weak $L^2$ topology as $n\to\infty$.
 
{\bf Proof:}
By a standard result in functional analysis {\ref}, weak $L^2$ convergence 
(that $\fh_n-\gh$ integrated against any $L^2$ test function approaches zero) 
follows from
the pointwise
convergence (Lemma C3) and bounded norms (Lemma C2).

{\bf Lemma C5:}
$||\fh_n||_2 \to ||\gh||_2$ as $n\to\infty$.

{\bf Proof:}
Letting $p\uparrow 1$ in Eq. $(C2)$
and invoking Fatou's Lemma,
$$\limsup_{n\to\infty} ||\fh_n||_2 \leq  ||\gh||_2.$$
But since $\fh_n\to\gh$ in weak $L^2$ topology, we have
$$||\gh||_2\leq \liminf_{n\to\infty} ||\fh_n||_2.$$
The two
preceding inequalities imply
$\limsup||\fh_n||_2\le\liminf ||\fh_n||_2$.
Since the reverse is true always, 
$\liminf = \limsup$, which implies 
that $\lim ||\fh_n||_2$ exists and equals $||\gh||_2$.

{\bf Theorem II:}
If $f$ is a proper quasi-probability density, then 
$$\int |f_n(\vx) - g(\vx)|^2 d^dx\to 0\as n\to\infty,$$
{\ie} $f_n$ approaches a Gaussian in $L^2$ norm.

{\bf Proof:}
Because of the Plancherel Theorem ($L^2$-unitarity of the Fourier transform),
this is equivalent to 
$$||\fh_n - \gh||_2 \to 0\as n\to\infty,$$
{\ie} that $\fh_n\to\gh$ in strong $L^2$ topology. 
But by a standard functional analysis result, this follows from 
weak $L^2$ convergence (Lemma C4) combined with convergence of the norm (Lemma
C5), so the proof is complete.

Thus we have shown that $f_n$ approaches Gaussianity in the strong $L^2$ sense. 
Note that in Lemma 1, we used the technical assumption that $f$ was either
integrable or a Wigner function. 
If we wish to make the additional 
technical assumption that not only is $f$ (and hence $\fh$) in 
$L^2$ but, for some $\epsilon>0$ (however small) 
$|\fh|^2 |\vk|^{\epsilon}$ is also
integrable over $\R^d$, then we can show the following:
$\fh_n$ converges not merely in strong $L^2$ but also in strong $L^1$, and
consequently  $f_n$
converges uniformly to a Gaussian. Thus 
$$\sup_{\vx} |f_n(\vx) - g(\vx)| \to 0\as n\to\infty,$$
which rules out 
a number of physically uninteresting
pathological cases, such as
$f_n(\vx)$ converging to the Gaussian
$g(\vx)$ for all $\vx$ except for a set of measure zero.

{\bf Lemma C6:}
If $W$ is a Wigner function, then
$\Wh(\vz) \leq \Wh(0) = 1$, where the inequality is strict if $\vz\neq
0$.

{\bf Proof:}
For a normalized wavefunction $\psi$ in $n$ dimensions (a pure state), 
the Fourier transform of the
Wigner function is
$$\Wh(\vk,\vx) = 
\int e^{-i(\vk\cdot\vq + \vx\cdot\vp)} W(\vq,\vp)\,d^nq\,d^np
= \int e^{-i\vk\cdot\vq}  
\psi(\vq-\vx/2)^* \psi(\vq+\vx/2)\,d^nq,$$
where the integral is to be taken over
all space. Thus
$$|\Wh(\vk,\vx)| \leq
\int
|\psi(\vq-\vx/2)|\>| \psi(\vq+\vx/2)|d^n q.$$
Using the trivial inequality $AB \leq (A^2+B^2)/2$ (with strict inequality 
unless $A=B$),
we obtain 
$$\eqalign{ |\Wh(\vk,\vx)| &\leq
{1\over 2}\int
|\psi(\vq-\vx/2)|^2 d^n q
+  
{1\over 2}\int
|\psi(\vq+\vx/2)|^2
 d^n q\crr
&= \int|\psi(\vq)|^2 d^n q = 1.
}$$
The case of a mixed state, where the Wigner function is a weighted average 
of Wigner functions of pure states, follows directly from superposition.

That we have strict inequality for $\vz\neq 0$ is seen as follows:
The second inequality above is an equality only if $\psi(\vq-\vx/2) =
\psi(\vq+\vx/2)$ almost everywhere, {\ie} if $\psi$ has period $\vx$. But since 
$\int \psi(\vq) d^n q = 1$, $\psi$ cannot be periodic, and the only possibility
is $\vx=0$.
Thus setting $\vx=0$ in the first inequality and subtracting unity from both sides
shows that we have equality only if
$$\int\left[1 -\cos(\vk\cdot\vq)\right]|\psi(\vq)|^2 d^nq =  0.$$
Since the integrand is non-negative, it must vanish identically. Since 
$||\psi||_2=1>0$, $\psi$ cannot vanish almost everywhere, and the only
possibility is $\vk=0$.

{\bf Theorem III:}
The Gaussian result in Theorem (II) is always obtained if $f$ is a Wigner
function with finite first and second moments. 

{\bf Proof:}
Let $f(\vz) = W(\vz) = W(\vp,\vq)$ be a Wigner-function in $n$ dimensions, {\ie}
take $d=2n$.
We only need to check that all the conditions of Theorem (II) hold, {\ie} that
all such Wigner functions are indeed proper quasi-probability densities.
All Wigner functions integrate to unity and are square-integrable (indeed 
$||W||_2 \leq (2\pi\hbar)^{-n/2}$, with equality only for pure states 
$[\number\HillaryClinton]$).
Lemma C6 showed that all Wigner functions are indeed proper. 
Thus all that remains to be shown is that 
the covariance matrix $V=C$ is strictly
positive definite. This is a well-know fact, basically a corollary to the 
multidimensional uncertainty principle uncertainty, but we give a brief 
proof here for completeness. 

Without loss of generality, we may
assume that the Wigner function corresponds to a 
pure state, since the covariance
matrix of a mixed state is simply the weighted average of the covariance
matrices in the mixture,  and the weighted average of positive definite matrices
is always positive definite.
That $V$ is positive semidefinite follows immediately from the fact that Wigner
functions are proper:
$$|\Wh(\vz)|^2 = 1 - V_{jk}\vz_j \vz_k + O(|\vz|^3),$$
where $j$ and $k$ are to be summed over from $1$ to $d$,
so if $V$ would have a negative eigenvalue, then there must exist a point near
the origin where $|\Wh|>1$, a contradiction. 
The multidimensional uncertainty relationship 
$[\number\refnr,\nextref\number\refnr\nextref]$ 
states that 
$$\det V \ge (\hbar/2)^{2n}$$
for all Wigner functions, with equality only for Gaussian pure
states, so none of the eigenvalues of $V$ can vanish, and $V$ must be positive
definite. Thus all Wigner functions are proper
quasi-probability densities, and the proof is complete.

Note that in contrast to the case of positive densities, second
moments can vanish not merely in pathological cases,
but also for well-behaved functions.
Such an example is 
$$\Wh(\vz) = e^{-z^4},$$
for which $V=0$.
Also note that the requirement that the first and second 
moments be finite is necessary for
the classical CLT as well. Finite first and second moments with respect to
momentum is equivalent to the kinetic energy being finite. Finite first and
second spatial $\vq$-moments can be interpreted as the system
being spatially localized. Indeed, if the Hamiltonian is quadratic and positive
definite (as it was in all cases treated in this paper), then all first and
second moments must be finite if the total energy of the system is finite.

\vfill
\break
\baselineskip12pt
\frenchspacing

\beginsection  REFERENCES

\smallskip
\refnr=1

\rf W. H. Zurek, S. Habib \& J. P. Paz (``ZHP");Phys. Rev. Lett.;70;1187;1993


\rf O. K\"ubler \& H. D. Zeh;Ann. Phys.;76;405;1973

\rf W. H. Zurek;Phys. Today;44 (10);36;1991

\rf E. Joos \& H. D. Zeh;Z. Phys. B;59;223;1985

\rf M. R. Gallis \& G. N. Fleming;Phys. Rev. A;42;38;1989

\rf M. Tegmark;Found. Phys. Lett;6;571;1993

\rn{L. Di\'osi, in {\it `Proc. of the Workshop in Trani on Waves and Particles in
Light and Matter}, eds. A. Garuccio, A. vd der Merwe (Kluwer, 1993).}

\rf S. W. Hawking;Commun. Math. Phys.;87;395;1982

\rf J. Ellis, S. Mohanty \& D. V. Nanopoulos;Phys. Lett. B;221;113;1989


\rf M. A. Huerta and H. S. Robertson;J. Stat. Phys.;1;393;1969

\rf Tegmark, M. \& Yeh, L. (``TY"); Physica A;202;342;1994

\rf G. W. Ford, J. T. Lewis \& R. F. O'Connell;J. Stat. Phys.;53;439;1988


\rf E. P. Wigner;Phys. Rev.;40;749;1932

\rf M. Hillery, R. H. O'Connell, M. O. Scully, \&
E. P . Wigner;Phys. Rep.;106;121;1984
 
\rn{Y. S. Kim and M. E. Noz, {\it Phase Space Picture of
Quantum Mechanics: Group Theoretical Approach} (World
Scientific, Singapore, 1991).}

\rf E. Schr\"odinger;Naturwissenschaften;14;664;1926

\rf R. J. Glauber;Phys. Rev.;131;2766;1963


\rn{P. J. Davis, {\it Circulant Matrices} (Wiley, New York, 1979).}

\rf G. W. Ford, M. Kac \& P. Mazur;J. Math. Phys.;6;504;1965
 

\rn{A. Renyi, {\it Foundations of Probability}
(Holden-Day, San Francisco, 1970).}


\rf W. Hoffding \& H. Robbins;Duke Math. J.;15;773;1948

\rf Z. A. Lomnicki \& S. K. Zaremba;Math. Zeitschr.;66;490;1957


\rn{W. Feller, {\it An introduction to probability theory and 
its applications, 3rd ed.} (Wiley, New York,1968).}

\rn{Prohorov, Y. V. \& Rozanov, Y. A.,
{\it Probability Theory} (Springer, Berlin, 1969).}


\rf E. Schr\"odinger;Ann. Phys. (Leipzig);44;916;1914

\rf P. Mazur \& E. Montroll; J. Math. Phys;1;70;1960


\rn{R. P. Feynman, {\it Statistical Mechanics}
  (Benjamin, Reading, 1972).}


\rn{A. Zygmund, {\it Trigonometric Series}, p.198 
(Campridge Univ Press, Cambridge, 1968).}

\rn{E. Hewitt \& K. Stromberg, 
{\it Real and Abstract Analysis} (Springer, Berlin 1965).}


\rf R. Simon, E. C. G. Sudarshan \& N. Mukunda;Phys. Lett.;124A;223;1987

\rn{L. Yeh, Ph.D. Thesis (U.C. Berkeley 1993).}

\end